\documentclass[10pt]{article}
\usepackage[T1]{fontenc}
\usepackage[utf8]{inputenc}
\usepackage{authblk}
\usepackage{url}
\usepackage{xcolor}
\usepackage[lofdepth,lotdepth]{subfig}
\usepackage{setspace}
\usepackage{multicol}
\usepackage{amsmath}
\usepackage{multirow}
\usepackage{booktabs}
\usepackage{wrapfig}
\usepackage[letterpaper, top=1in, bottom=1in, left=1in, right=1in,headheight=30pt]{geometry}
\usepackage{natbib}
\usepackage{graphicx}
\usepackage{float}
\usepackage{mathptmx}
\usepackage[section]{placeins}
\usepackage{nameref}
\usepackage{dblfloatfix}
\usepackage{fixltx2e}
\usepackage{etoolbox}
\usepackage{sectsty}
\usepackage{ulem}
\usepackage[labelfont=bf]{caption}
\usepackage{csquotes}

\usepackage{fancyhdr}
\usepackage{bm}
\usepackage{amssymb}
\usepackage{dsfont}

\makeatletter
\let\ps@plain\ps@fancy 
\makeatother

\makeatletter
\patchcmd{\@maketitle}{\LARGE \@title}{\fontsize{18}{19.2}\selectfont\@title}{}{}
\makeatother

\makeatletter
\renewcommand\AB@affilsepx{, \protect\Affilfont}
\makeatother

\renewcommand\Affilfont{\fontsize{14}{0}}

\sectionfont{\fontsize{12}{0}\selectfont}
\subsectionfont{\fontsize{12}{0}\selectfont}
\subsubsectionfont{\fontsize{10}{0}\normalfont\underline}

\begin{document}
%\pagestyle{plain}
%\pagenumbering{gobble}
\setcitestyle{round}
%Title Page

\title{\bf Reduced Order Model of a Generic Submarine for Maneuvering Near the Surface }
\author[1]{\vspace{-12pt}J.~Ezequiel~Martin}
\author[2]{Maxwell~Hammond}
\author[2]{Nicholas~Rober}
\author[1]{Yakin~Kim}
\author[1,2]{Venanzio~Cichella}
\author[1]{Pablo~Carrica}
\affil[ ]{(\textsuperscript{$1$}IIHR - Hydroscience and Engineering, The University of Iowa, USA, \textsuperscript{$2$}Department of Mechanical Engineering, The University of Iowa, USA)}
% \affil[1]{\vspace{-3pt}IIHR - Hydroscience and Engineering, The University of Iowa, USA}
% \affil[2]{\vspace{-3pt}Department of Mechanical Engineering, The University of Iowa, USA}

\renewcommand\Authands{ and }
\date{}

\maketitle
\thispagestyle{fancy}
\begin{multicols}{2}

\section*{ABSTRACT}

A reduced order model of a generic submarine is presented. Computational fluid dynamics (CFD) results are used to create and validate a model that includes depth dependence and the effect of waves on the craft. The model and the procedure to obtain its coefficients are discussed, and examples of the data used to obtain the model coefficients are presented. An example of operation following a complex path is presented and results from the reduced order model are compared to those from an equivalent CFD calculation. The controller implemented to complete these maneuvers is also presented.  

\section*{INTRODUCTION}

Reduced order models (ROM) are a suitable option for the development of control algorithms and path planning for submarines. ROMs represent a trade-off between accuracy and execution cost but, if able to correctly reproduce the dynamic response of the vehicle, are an excellent tool to study maneuvering under external disturbances, such as waves \citep{fang2006wave}, or for collision avoidance \citep{jin2020dynamic}, in particular if multiple simulations are required for optimization. The ROM used in this work has been developed explicitly to consider wave and depth effects and to support the development of advanced controllers \citep{rober2021three}.

Modeling operation near the surface, in restricted waters, and at low speeds is a particularly challenging problem as craft controllability becomes compromised by the reduced authority of its control surfaces or more demanding requirements are present as danger
of collision or surfacing increase. The development of controllers and control strategies is greatly aided by dynamic models of the craft that are fast, yet accurate. While experimental or high-fidelity computations can also be used for testing, they are considerably more expensive and time consuming. Early development of control strategies and algorithms and subsequent initial tuning are typically performed using ROM, leaving experiments and CFD for fine tuning or to study controller performance under more complex effects not captured by simpler ROM approaches. 

The ROM solves the rigid body equations of motion of the vehicle under external forces and moments. The external forces in a submarine include the hydrostatic and hydrodynamic forces. The hydrodynamic forces are a result of the state of motion of the vessel, in the form of virtual mass, pressure drag, and skin friction; the objective of the ROM is to relate those forces to the kinematics of the craft. Accurate prediction of the hydrodynamic loads requires information under wide motion conditions and the calibration of a large number of coefficients. The most frequently used form of hydrodynamic model (HDM) for cruciform stern plane configurations was proposed by \cite{gertler1967standard}. In this HDM only forces and moments due to certain motions are considered and the history of motions is neglected. The coefficients or derivatives required by the model are evaluated from experiments, semi-empirical approaches, or computational methods.

Experimental methods are the most conventional technique to obtain the coefficients by performing captive model tests with scaled models. Towing with or without Planar Motion Mechanisms (PMM) and Rotating Arms (RA) are used to isolate the desired coefficients \citep{feldman1995method}. For submarines, wind tunnel tests are also possible for deep conditions. The validation of HDM can be done experimentally by performing free running controlled maneuvers in model scale in a wave basin \citep{overpelt2015free}.  Experimental techniques provide reliable data but require expensive facilities and construction of a model. In addition, experiments are limited to model scale and require extrapolation of results  to full scale.

Computational techniques can also be used for coefficient evaluation. Potential flow solvers are cost-effective tools to estimate pressure effects and virtual mass. However, this methodology is not accurate for separated flows, providing poor estimates at large angles of attack \citep{evans2004dynamics}. CFD can also be used for this purpose, however the simulation cost may limit its use to relatively simple geometries or require the use of very coarse grids. In the past, unsteady tests such as PMM might have been replaced by static calculations to reduce the computational requirements. Current capabilities have allowed the simulation of more challenging conditions, and full maneuvers are routinely completed using moving control surfaces \citep{carrica2021cfd}. The success in predicting maneuvers with CFD lends credibility to its use to obtain the coefficients of ROM, but development of accurate ROM simulation models is a significant challenge as the operational conditions deviate from those used to obtain the coefficients.

A reduced order model  for the generic submarine Joubert BB2 is presented.  The model originally proposed by \cite{gertler1967standard} has been updated to include a X-shape for the stern planes and a simplified formulation of some of its terms. More importantly, surface effects and added mass effects due to wave action have been included in the model, allowing the simulation of maneuvers near the surface. Extensive computational fluid dynamics  tests were conducted using REX, a CFD solver developed at The University of Iowa, to generate the model coefficients. The resulting model is implemented in the commercial software MATLAB SIMULINK\texttrademark  with the purpose of generating an open model for a variety of applications. The resulting hydrodynamic model (HDM) has been satisfactorily compared to available experimental and numerical data for Joubert BB2 \citep{carrica2019,carrica2021cfd} and used to support the development of a novel controller \citep{rober2021three}. The model provides a quick platform for evaluation of control strategies, planning of maneuvers, etc. As an example of the capabilities of the ROM and the controller, a 6DoF path following maneuver simulated using both the ROM and CFD calculation are presented.  
The SIMULINK model along with tutorials and inputs for the validation case presented in this paper and others can be found at our github web page\footnote{\url{https://github.com/caslabuiowa/IowaBB2model}}. 

\section*{METHODS}
\subsection*{Joubert BB2}
The geometry used in this work is based on the Joubert  \citep{joubert2004some,joubert2006some} originally developed by the Australian DSTO and further enhanced by MARIN in The Netherlands \citep{overpelt2015free}. The main particulars of both model and prototype are presented in Table~\ref{tab:BB2}. The model includes a X-shape arrangement of control stern planes and a set of sail planes. The original control of BB2 consisted of a  pair of PD controllers for direction, that combine control of depth and pitch in a vertical command $\delta_V$ and of lateral displacement and yaw in a horizontal command $\delta_H$. The control planes are then actuated as combination of the two commands:
\begin{equation}
\begin{split}
    \delta_1&= -\delta_H - \delta_V \\
    \delta_2&= \delta_H - \delta_V \\
    \delta_3&= \delta_H + \delta_V \\
    \delta_4&= -\delta_H + \delta_V \\
    \delta_5&= - \delta_V \\
    \end{split}
    \label{CommandsVH}
\end{equation}
where $\delta_1$ is the lower starboard plane, $\delta_2$ the upper starboard, $\delta_3$ the upper port, $\delta_4$ the lower port and $\delta_5$ are the sail planes. Additional control options such as ballast and trim tanks have been explored numerically \citep{carrica2019} and research in advanced controllers to improve off-design operation is on-going \citep{rober2021three}.
\begin{table}[H]
    \centering
    \caption{Main particulars of Joubert BB2}
    \begin{tabular}{l l c c}
\toprule
&\textbf{Symbol}&\textbf{Full}&\textbf{Model}  \\
\midrule
\!Length&$L_0$ (m)&70.2 & 3.16\\
\!Beam&$B$ (m)&9.6&0.5232\\
\!Depth to top of sail &$D_0$  (m)&16.2&0.8829\\
\!Displacement&$\nabla$ (tons)&4440&0.7012\\
\!Center of Gravity \\ \quad Long. (from nose)&$X_G$  (m)&32.31&1.761\\
\quad Vertical  (from shaft)&$Z_G$  (m)&0.0443&0.0024\\
\!Gyration radii\\ \quad Roll&$r_x$  (m)&3.433&0.1871\\
\quad Pitch &$r_y$  (m)&17.6&0.9592\\
\quad Yaw &$r_z$  (m)&17.522&0.955\\
\bottomrule

    \end{tabular}
    \label{tab:BB2}
\end{table}

\subsection*{Hydrodynamic Model}\label{ExpModel}
The ROM solves the six degrees of freedom (6DoF) equations of motion of the craft in a coordinate system local to the body: 
\begin{equation} \label{eq:sixdofgen} 
M \dot{s} = F-b=F_b+F_h-b,  
\end{equation} 
where $s= [u,v,w,p,q,r]^T$ is the generalized velocity vector, $M$ is the mass matrix, given as a function of the mass $m$, the center of gravity of the craft $(x_G,y_G,z_G)$ and the inertia tensor $I$ as: 
\begin{equation} \label{eq:M}
\footnotesize{
 	\!M\!=\!\begin{bmatrix} m&0 & 0 & 0 &m z_G & -my_G \\ 
 	                0&m & 0 & -mz_G &0 & mx_G \\ 
 	                0&0 & m & my_G &-mx_G & 0 \\ 
 	                0&-mz_G & my_G &I_{xx} &-I_{xy} & -I_{xz} \\ 
 	                mz_G&0 & -mx_G &-I_{xy} &I_{yy} & -I_{yz} \\ 
 	                -my_G&mx_G & 0 &-I_{xz} &-I_{yz} & I_{zz} \\ 
 	                \end{bmatrix}\!}
\end{equation}

\begin{figure}[H]
  \centering
    \includegraphics[width=\columnwidth]{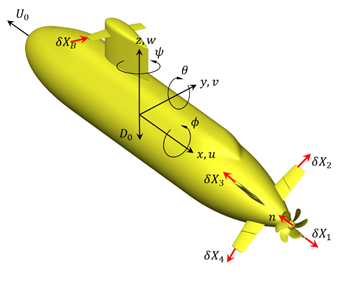}
    \caption{Definition of axes and variables used for modelling BB2.}
    \label{fig:axis_def}
\end{figure}

Since the equations are solved in the ship system (shown in Fig.~\ref{fig:axis_def}), all extra diagonal terms are typically zero but are included in the general model to consider changes to the mass distribution through actuation of trim tanks for control. 
 The coupling terms due to the use of a non-inertial reference system, $b$ is 
\begin{equation} \label{eq:b}
\small{
 b=	\begin{bmatrix} m[wq\!-\!vr\!-\!x_G (q^2\!+\!r^2 )\!+\!y_G (pq\!-\!\dot{r} )\!+\!z_G (pr\!+\!\dot{q} )]\\
 	                m[ur\!-\!wp\!-\!y_G (r^2\!+\!p^2 )\!+\!z_G (qr\!-\!\dot{p} )\!+\!x_G (qp\!+\!\dot{r} )]\\
 	                m[vp\!-\!uq\!-\!z_G (p^2\!+\!q^2 )\!+\!x_G (rp\!-\!\dot{q} )\!+\!y_G (rq\!+\!\dot{p} )]\\
 	                  (I_{zz}\!-\!I_{yy})qr\!+\!m[y_G (\dot{w}\!-\!uq\!+\!vp)\!-\!z_G (\dot{v}\!-\!wp\!+\!ur)]\\
 	                  (I_{xx}\!-\!I_{zz})rp\!+\!m[z_G (\dot{u}\!-\!vr\!+\!wq)\!-\!x_G (\dot{w}\!-\!uq\!+\!vp)]\\
 	                  (I_{yy}\!-\!I_{xx} )pq\!+\!m[x_G (\dot{v}\!-\!wp\!+\!ur)\!-\!y_G (\dot{u}\!-\!vr\!+\!wq)] \\
 	                  \end{bmatrix}}
\end{equation}
The external loads $F$ are split in hydrostatic $F_b$ and hydrodynamic forces $F_h$. Craft weight $W$ and buoyancy $B$ are included in $F_b$, considering the possibility of changes to these terms due to control mechanisms that change total mass of the craft (ballast tank), or its distribution (trim tanks):
\begin{equation} \label{eq:F_b}
\small{ 	F_b \!=\!
 	\begin{bmatrix}
 	-(W-B)  \sin{\theta}\\
 	  (W-B)  \cos{\theta}  \sin{\phi}\\
 	  (W-B)  \cos{\theta} \cos{\phi} \\
 	  (y_G W\!-\!y_B B)\cos{\theta}\cos{\phi}\!-\!(z_G W\!-\!z_B B)\cos{\theta} \sin{\phi}\\
 	  (z_G W\!-\!z_B B) \sin{\theta}\!-\!(x_G W\!-\!x_B B)\cos{\theta}\cos{\phi}\\
 	  (y_G W\!-\!y_B B) \sin{\theta}\!-\!(x_G W\!-\!x_B B)\cos{\theta}\sin{\phi}\\	                  \end{bmatrix}}
\end{equation}
The external load $F_h$
 parameterizes the loads  acting on the craft in terms of relevant kinematic variables such as craft speed, and angle of attack (AoA) of the craft and its control surfaces, as well as external parameters, including distance to the surface and sea state.   The general form of the six forces and moments $F_{h,i} $ is

\begin{equation} \label{eq:sixdof_fh} 
\begin{split}
F_{h,i} =\sum_{k=1}^6\sum_{j=1}^6 F'_{i,s_j s_k} s_j s_k + \sum_{j=1}^6 F'_{i,s_j} \dot{s}_j +\\ \sum_{l=1}^5 F'_{i,\delta X_l} u^2 \delta X_l^2+F_{i,prop}.
\end{split}
\end{equation} 

The hydrodynamic loads are grouped in four terms. The first group includes drag terms and added mass effects due to motions. The second group corresponds to added mass due to acceleration. Virtual mass and wave effects are considered by discretizing the boat geometry into sections to generate lumped coefficients for each section. The third group corresponds to the forces and moments generated by the effective deflection 
$\delta X_l$ of each control surface (four stern planes in cruciform shape, plus the sail plane); and finally the propeller forces and moments  are also considered.  

Many of the terms in this general form are shown to be negligible compared to the dominant ones and are not included in the final model. Slight modifications to the general form presented in Eq.~\eqref{eq:sixdof_fh}, such as the lumping of lateral forces or the use of absolute value instead of the signed value of some variables is used to further simplify the model. A  description of the terms and the different numerical simulations used to obtain them is presented in the next section.  The longitudinal force $F_{h,1} = X_h$ is approximated by
\begin{equation} \label{eq:fh_x} 
\begin{split}
X_h = \frac{\rho L^2}{2}  \bigg[X'_{uvw}(U,\alpha,\beta,z) U^2 +\sum_{l=1}^5 X'_{\delta X_l} u^2 \delta X_l^2 \bigg]+\\
\frac{\rho L^3}{2}   \bigg[ X'_{\dot{u}}\dot{u} + X'_{vr}vr+X'_{wq} wq\bigg] + \\
\frac{\rho L^4}{2}   \bigg[X'_{qq}q^2+X'_{rr} r^2 +X'_{rp} rp\bigg]+ X_{prop}.
\end{split}
\end{equation} 
In Eq.~\eqref{eq:fh_x}, $\alpha$ and $\beta$ are respectively the drift and incidence angle, $U$ the velocity magnitude and $X_{prop}$ the propeller thrust. The other forces and moments are as follows:
\begin{equation} \label{eq:fh_y} 
\begin{split}
Y_h = \frac{\rho L^2}{2}\bigg[Y'_{uvw}(U,\alpha,\beta,z) U^2 +\sum_{l=1}^5 Y'_{\delta X_l} u^2 \delta X_l^2\bigg]+\\
\frac{\rho L^3}{2}   \bigg[ Y'_{\dot{v}}\dot{v}+ Y'_{up}up+ Y'_{ur}ur+ Y'_{vq}vq + Y'_{wp}wp+  \\  Y'_{wr}wr + Y'_{v|r|}\frac{v}{|v|}(v^2+w^2)^{1/2}|r| \bigg] + \frac{\rho L^4}{2}   \bigg[Y'_{\dot{r}}\dot{r}  +\\  Y'_{\dot{p}}\dot{p}+Y'_{|p|p}|p|p+Y'_{pq}pq+Y'_{pr}pr \bigg] +
Y_{prop};
\end{split}
\end{equation} 

\begin{equation} \label{eq:fh_z} 
\begin{split}
Z_h = \frac{\rho L^2}{2}\bigg[Z'_{uvw}(U,\alpha,\beta,z) U^2 +\sum_{l=1}^5 Z'_{\delta X_l} u^2 \delta X_l^2\bigg]+\\
\frac{\rho L^3}{2}   \bigg[ Z'_{\dot{w}}\dot{w}+ Z'_{vp}vp+ Z'_{vr}vr+ Z'_{uq}uq +\\  Z'_{w|q|}\frac{w}{|w|}(v^2+w^2)^{1/2}|q| \bigg] +\\ \frac{\rho L^4}{2}   \bigg[Z'_{\dot{q}}\dot{q} +Z'_{pp}p^2+Z'_{rr}r^2+Z'_{pr}pr \bigg] +
Z_{prop};
\end{split}
\end{equation} 

\begin{equation} \label{eq:fh_p} 
\begin{split}
K_h \!=\! \frac{\rho L^2}{3}\bigg[K'_{uvw}(U,\alpha,\beta,z) U^2 \!+\!\sum_{l=1}^5 K'_{\delta X_l} u^2 \delta X_l^2\bigg]+\\
\frac{\rho L^4}{2}   \bigg[K'_{\dot{v}}\dot{v}+ K'_{up}up+ K'_{ur}ur+ K'_{vq}vq + K'_{wp}wp+ \\ K'_{wr}wr  \bigg] + \frac{\rho L^5}{2}   \bigg[K'_{\dot{p}}\dot{p} +K'_{r}\dot{r}+K'_{qr}qr+\\K'_{pq}pq+K'_{|p|p}|p|p \bigg] +
K_{prop};
\end{split}
\end{equation} 
\begin{equation} \label{eq:fh_r} 
\begin{split}
M_h = \frac{\rho L^2}{3}\bigg[M'_{uvw}(U,\alpha,\beta,z) U^2 +\sum_{l=1}^5 M'_{\delta X_l} u^2 \delta X_l^2\bigg]\\+
\frac{\rho L^4}{2}   \bigg[M'_{\dot{w}}\dot{w}+ M'_{uq}uq+ M'_{vp}vp + M'_{vr}vr+ \\ M'_{|vw|q}q (v^2+w^2)^{1/2} \bigg] + \frac{\rho L^5}{2}   \bigg[M'_{\dot{q}}\dot{q} +M'_{pp}p^2\\+M'_{rr}r^2+M'_{rp}rp+M'_{|q|q}|q|q \bigg] +
M_{prop};
\end{split}
\end{equation} 
\begin{equation} \label{eq:fh_q} 
\begin{split}
N_h\!=\!\frac{\rho L^2}{3}\bigg[N'_{uvw}(U,\alpha,\beta,z) U^2\!+\!\sum_{l=1}^5 N'_{\delta X_l} u^2 \delta X_l^2\bigg]+\\
\frac{\rho L^4}{2}   \bigg[N'_{\dot{v}}\dot{v}+ N'_{up}up+ N'_{ur}ur+ N'_{vq}vq + N'_{wp}wp+ \\  N'_{|vw|r}r (v^2+w^2)^{1/2} \bigg] + \frac{\rho L^5}{2}   \bigg[N'_{\dot{r}}\dot{r} +N'_{p}\dot{p}+\\N'_{pq}pq+N'_{qr}qr+N'_{|r|r}|r|r \bigg] +
N_{prop}.
\end{split}
\end{equation} 
Some of these terms are currently not included in the model as it continues to be developed, however the validations conducted show that the model is adequate for the purpose it was intended, namely the development of controllers, and the missing terms are non dominant in the craft dynamic response.

\begin{figure}[H]
  \centering
    \includegraphics[width=\columnwidth]{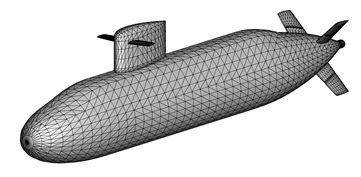}
    \caption{Triangulated surface for hydrostatic load calculation. The total number of elements $N_e$ is 7,148.}
    \label{fig:hydrostat}
\end{figure}

Wave-induced loads are incorporated to the model by integration over a coarse grid as shown in Fig.~\ref{fig:hydrostat}. While the wave pressure force is dynamic, it is herein considered independent from the state of the vehicle and thus only a function of time and space, allowing grouping with the hydrostatic force. For fully submerged, neutrally buoyant conditions, the weight is adjusted to exactly match the buoyancy obtained by integration as initial condition, and allowed to evolve if a ballast controller is used, and restoring moments are calculated as shown in Eq.~\eqref{eq:F_b}. The additional forces and moments due to waves are calculated by integration of the pressure at the Gauss points of the grid shown in Fig.~\ref{fig:hydrostat} as 

\begin{equation} \label{eq:hydrostat_mom} 
\begin{split}
F_{w,i} &=-\frac{1}{3}\sum_{i=1}^{N_e}(p_{12,i}+p_{23,i}+p_{31,i})A_i;\\
M_{w,i}\! &=\!-\frac{1}{3}\sum_{i=1}^{N_e}(r_{12,i}p_{12,i}\!+\!r_{23,i}p_{23,i}\!+\!r_{31,i}p_{31,i})\!\times\! A_i;\\
\end{split}
\end{equation} 
with $A_i$ the area vector of each element and $r_{i}$ the coordinates of the Gauss point in the ship system. The pressure field is given analytically as that of a progressive regular wave in deep water

\begin{equation} \label{eq:wave} 
\begin{split}
p\! =\!-\rho g z + \rho g A_{wave} e^{k_{wave}z} \sin(k_{wave} x - \omega_{wave} t)
\end{split}
\end{equation} 
where $k_{wave}$ is the wave number, $\omega_{wave}$ is the wave frequency in rad/s, $A_{wave}$ is the amplitude and $z$ is the vertical distance to the calm water level. This method can be extended to include more complex wave fields.  

\subsection*{Computational Fluid Dynamics}\label{CFD}

 The coefficients in Eqs.~\eqref{eq:fh_x} through \eqref{eq:fh_q} need to be modeled from existing numerical or experimental data. In the present work, high-fidelity computational fluid dynamics results, generated with REX is used. REX solutions for BB2 have been extensively validated against experimental data \citep{carrica2021cfd}.
 REX is a  hybrid RANS/LES solver based on the SST turbulence model \citep{menter1994two} with dynamic overset capabilities \citep{noack2009suggar}. The free surface, included in the calculations used to develop the HDM, is modelled using single-phase level set \citep{carrica2007ship}. Further details on REX capabilities and numerical implementation can be found in \citep{li2020modeling}.
 
 A coarse multiblock mesh was used for coefficient calculation and for validation runs. The mesh allows the motion of discretized control surfaces, but does not include a discretized propeller, which is instead replaced by a body force. The coarse grid (4.5 M grid points) has been used in the past for validation purposes \citep{carrica2021vertical} and it is known to produce acceptable results. While a finer grid could be used to reduce errors in the coefficients estimates, it is expected that the main source of inaccuracies is in the ROM approach itself, and thus the additional numerical cost is not justifiable.

\subsection*{Model Coefficient Calculation}\label{ModelCoeff}

Specific simulations were conducted to estimate the coefficients in Eqs.~\eqref{eq:fh_x} to \eqref{eq:fh_q}. The simulations completed include self-propulsion calculation at different speeds and depths to compute thrust deduction coefficients; constant imposed acceleration to determined added mass coefficients; static drift computations; appendages forces for static conditions; and pure rotation and rotating arm simulations. Details for all of these computations are not presented herein, but instead a description of the results for some of these simulations is highlighted. The complete set of simulations and coefficient calculation is presented in \cite{kimthesis}.

Hull and sail resistance are dominant terms in the hydrodynamic load equation. Equally important are the forces and moments of the control surfaces to allow maneuvering. Properly characterizing these terms resulted in a large number of simulations in static conditions at multiple depths, towing speeds and angles of incidence and drift. Specifically, sail top depths $D_0$ = 2.5, 4.0, 7.0 and 25~m, and towing speeds $U_0$ = 3, 6 and 10~kts were used, and intervals of 4 degrees were considered for the drift angles in the range $-12\leq\alpha\leq12$ for vertical incidence and $0\leq\beta\leq12$ for horizontal drift, assuming lateral symmetry. Steady forces and moments for all solid surfaces are included in terms $F'_{i,uvw}$ in Eqs.~\eqref{eq:fh_x} to \eqref{eq:fh_q}, but excluding the contribution of the control surfaces. Simulations were conducted including the control surfaces at neutral position.

\begin{figure}[H]
  \centering
    \includegraphics[width= 5.in]{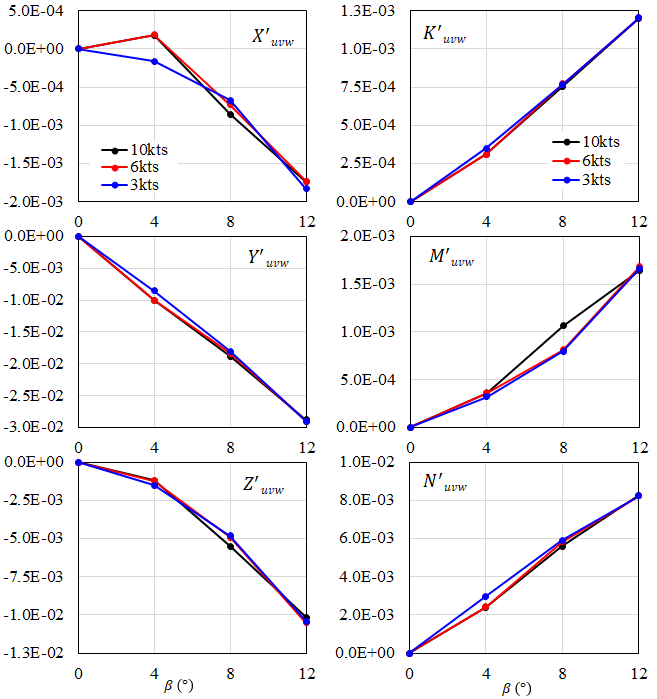}
    %%\captionsetup{width=6 in,indention=10pt}
    \caption{Drift force coefficients as a function of towing speed and horizontal drift angle $\beta$: left column forces coefficients $X'_{uvw}$, $Y'_{uvw}$ and $Z'_{uvw}$; right column moments coefficients $K'_{uvw}$, $M'_{uvw}$ and $N'_{uvw}$. All results at $D_0 = 25$ m. Values given as difference with straight ahead, even keel resistance.}
    \label{fig:hull_drift_moments_beta}
\end{figure}

\begin{figure}[H]
  \centering
    \includegraphics[width=2.6in]{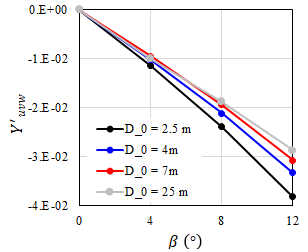}
    \caption{Lateral drift force coefficient as a function of  angle $\beta$ and depth $D_0$. All results at 10 kts.}
    \label{fig:hull_drift_y_depth}
\end{figure}

All six coefficients $F'_{i,uvw}$ as a function of $\beta$ and speed for $\alpha = 0$ and $D_0 = $25~m are shown in Fig.~\ref{fig:hull_drift_moments_beta}. As expected the lateral force is the dominant term, opposing the imposed motion. Very little velocity dependence is observed, and the force is proportional to the deflection. Also proportional to the drift angle are the yaw and roll moments; all other terms are smaller in magnitude and their behavior is approximately quadratic in the range considered. For all cases there are very little differences with towing speed indicating that the process is strongly pressure dominated. Depth of operation has a significant effect, as shown in Fig.~\ref{fig:hull_drift_y_depth}, with an increase of lateral force of about a third between the deep and shallowest case. The deepest condition, $D_0 = 25$~m, was considered representative of any deep conditions and constant coefficients were used for $D_0 \geq 25$~m; a similar conclusion cannot be made for operation outside the simulated range towards small depths, and results using the model for $D_0 \leq 2.5$~m should be considered unreliable.

\begin{figure}[H]
  \centering
    \includegraphics[width=5.6in]{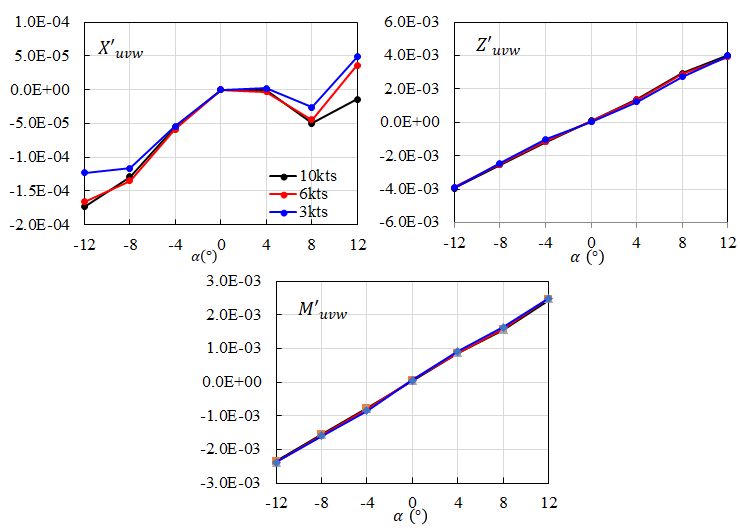}
    \caption{Drift force coefficients as a function of towing speed and angle $\alpha$ for forces coefficients $X'_{uvw}$  and $Z'_{uvw}$ and moment coefficients $M'_{uvw}$. All results at $D_0 = 25$~m. Values given as difference with straight ahead, even keel resistance.}
    \label{fig:hull_drift_alpha}
\end{figure}

Dependence on the  angle $\alpha$ is only shown for $X'_{uvw}$, $Z'_{uvw}$, and $M'_{uvw}$, since variations for the other coefficients are negligible. As before, only the difference with respect to the resistance in even keel conditions is included in Fig.~\ref{fig:hull_drift_alpha} While vertical force and pitch moment change linearly with the drift angle, the resistance $X'_{uvw}$ has a more complex variation with speed and incidence angle, particularly for $|\alpha|\geq8$. The presence of the sail structure makes the geometry significantly asymmetric and affects the longitudinal resistance due to differences in separated vortical structures. These differences are however small in the other two components compared to the overall effect of the hull in pitch and vertical force. Note that any difference is captured by the model since all available values are used for interpolation, rather than attempting to adjust a polynomial curve to the data.

Combined vertical and horizontal drift has not been included as an input to the model and effectively the forces are interpolated separately for each angle and then superimposed during simulation as independent processes. The inclusion of combined terms is a possible extension of the model to improve accuracy, but it is expected that particularly at small angles the current implementation yields acceptable results.

\begin{figure}[H]
  \centering
    \includegraphics[width=\columnwidth]{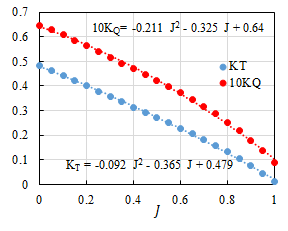}
    \caption{Open water thrust and torque coefficients $K_T$  and $K_Q$ for MARIN propeller 7371R.}
    \label{fig:POW}
\end{figure}

\begin{figure}[H]
  \centering
    \includegraphics[width=\columnwidth]{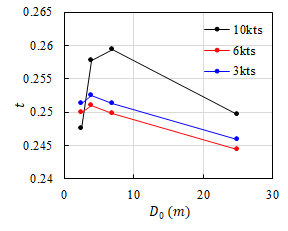}
    \caption{Thrust deduction at $D_0 =$ 2.5, 4, 7, and 25~m and $U_0$ = 3, 6, and 10 kts.}
    \label{fig:thrustdeduction}
\end{figure}

A second example of coefficient calculation is presented in Figs.~\ref{fig:POW} and \ref{fig:thrustdeduction} for the propeller performance. The open water performance curve for MARIN stock propeller 7371R, approximated as quadratic functions of the advance coefficient $J$ are used in the reduced order model directly to provide thrust and torque. The thrust deduction factor $t$ was evaluated from even-keel self-propulsion simulations at different depths and speeds.While the variations are small, clear trends of increasing $t$ towards the surface are observed up to $D_0 = $4~m. Near the surface decrease of $t$, particularly at large speeds due to an interaction with the free surface that limits the suction effect of the propeller.

The control surface performance was evaluated from even-keel constant speed CFD simulations, by imposing static deflection angles $\delta X_i$ within $-30^{\circ}\leq\delta X_i\leq30^\circ$ in intervals of $10^\circ$. Computations for different control planes were conducted simultaneously which can affect the results, however this is unavoidable unless the coefficients are evaluated for combinations of vertical and horizontal commands and the planes are deflected following the deflections prescribed by Eq.~\eqref{CommandsVH}.

\end{multicols}
\newpage

\begin{multicols}{2}

\begin{figure}[H]
  \centering
    \includegraphics[width=5.3in]{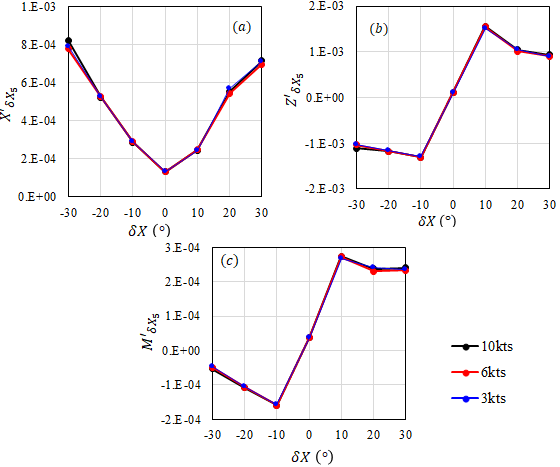}
    \caption{Sail plane coefficients of (a) resistance, (b) vertical force and (c) pitch moment for deflections$-30^{\circ}\leq\delta X_i\leq30^\circ$, at $U_0$ = 3 ,6, and 10 kts and $D_0$ = 25~m.}
    \label{fig:sailforces} 
\end{figure}

\begin{figure}[H]
  \centering
    \includegraphics[width=5.4in]{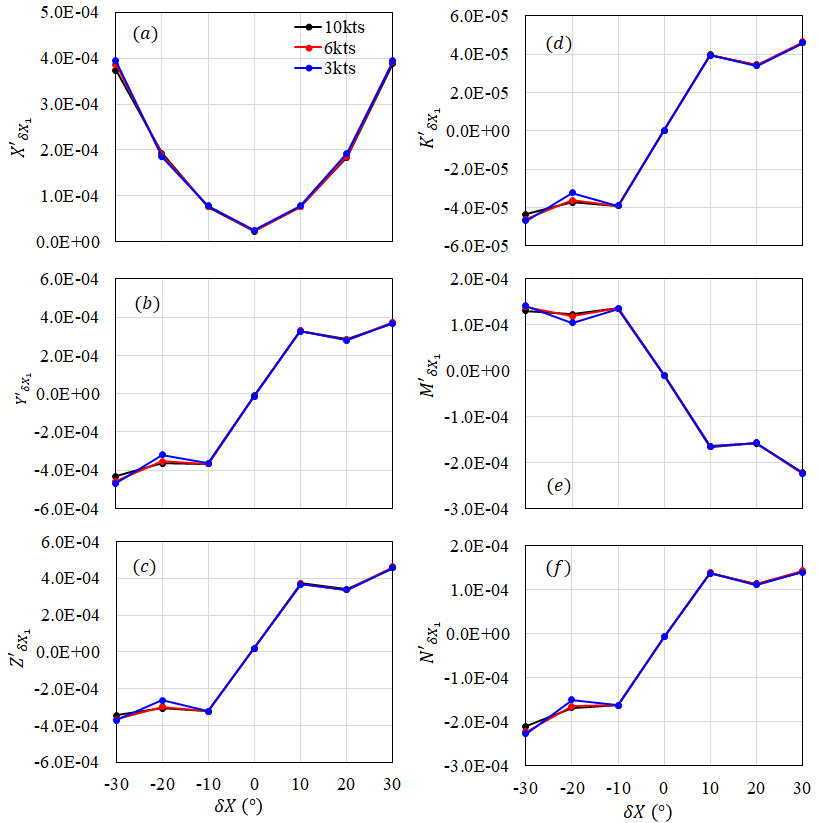}
    \caption{Lower starboard stern plane coefficients of (a) resistance, (b) lateral force, (c) vertical force, (d) roll moment, (e) pitch moment and (f) yaw moment for deflections $-30^{\circ}\leq\delta X_i\leq30^\circ$, at $U_0$ = 3, 6, and 10 kts and $D_0$ = 25~m.}
    \label{fig:SPForces}
\end{figure}

Figs.~\ref{fig:sailforces} and \ref{fig:SPForces} show control surface forces and moments at $D_0 = $25~m for the sail plane and the lower starboard stern plane, respectively. The sail and stern planes resistance increase non-linearly with the appendage deflection, while vertical and side forces and moments saturate as the fins stall at large angles of attack ($|\delta X_i |\leq 10^\circ$). The asymmetry of the sailplane location in the sail results in higher resistance when producing negative lift, as well as higher positive pitch moment.

Additional simulations used to characterized the model,  were presented by \cite{kimthesis}. These include the computation of virtual mass coefficients based on constant acceleration simulations; pure rotation simulations to obtain terms such as $X'_{qq}$ and $Y'_{|p|p}$; and RA simulations for coupled terms such as $X'_{uq}$. 

\end{multicols}

%\begin{figure}[H]
%  \centering
%    \includegraphics[width=2.8in]{Figures/hull_drift_forces_alpha.png}
%    \caption{Drift force coefficients as a function of towing speed and incidence angle %$\alpha$ for forces coefficients $X'_{uvw}$  and $Z'_{uvw}$and moment coefficients $M'_{uvw}$ %. All results at $D_0 = 25$ m. Values given as difference with straight ahead, even keel %resistance.}
%    \label{fig:hull_drift_alpha}
%\end{figure}

\begin{multicols}{2}

\end{multicols}
%\begin{figure}[H]
%  \centering
%    \includegraphics[width=\columnwidth]{Figures/SailForces.png}
%    \caption{Sail plane coefficients of (a) resistance, (b) vertical force and (c) pitch moment for deflections$-30^{\circ}\leq\delta X_i\leq30^\circ$, at $U_0$=3,6, and 10kts and $D_0$=25m.}
%    \label{fig:sailforces} 
%\end{figure}

\begin{multicols}{2}

\subsection*{Controller }\label{Controller}
\begin{figure}[H]
    \centering
    \includegraphics[width=\columnwidth]{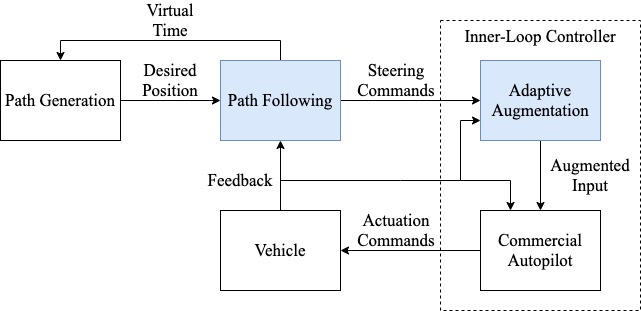}
    \caption{The overall control structure highlighting the path-following controller and the adaptive augmentation algorithm}
    \label{fig:pf_block_diagram}
\end{figure}

The Joubert BB2 reduced order model has been used as a basis for validation and testing of a multilayer control architecture comprised of motion planning, path following, and an inner-loop controller. This control strategy is shown in Fig.~\ref{fig:pf_block_diagram} and operates as follows: (1) the trajectory generation algorithm receives information about the vehicle and the environment to plan a feasible trajectory, (2) the computed trajectory is passed to the path-planning algorithm which generates kinematic steering commands, and (3) the inner-loop controller determines the control surface inputs necessary to execute the desired steering commands from the path-following controller. This controller has been developed considering the hazardous operating conditions autonomous underwater vehicles (AUVs) are often subject to in their missions, so its performance bounds have been rigorously considered to avoid potential obstacles or unwanted interaction with the surface.

The motion planning strategy employed in this architecture uses Bernstein polynomial approximation as a means of generating a continuous path, $\bm{p_d}(\cdot)$, from a set of defined control points. Optimization techniques demonstrated in \cite{cichella2019optimal} can be used to generate approximately optimal trajectories for the vehicle, while constraining limits to parameters like vehicle speed and angular rate.

\begin{figure}[H]
    \centering
    \includegraphics[width=.9\columnwidth]{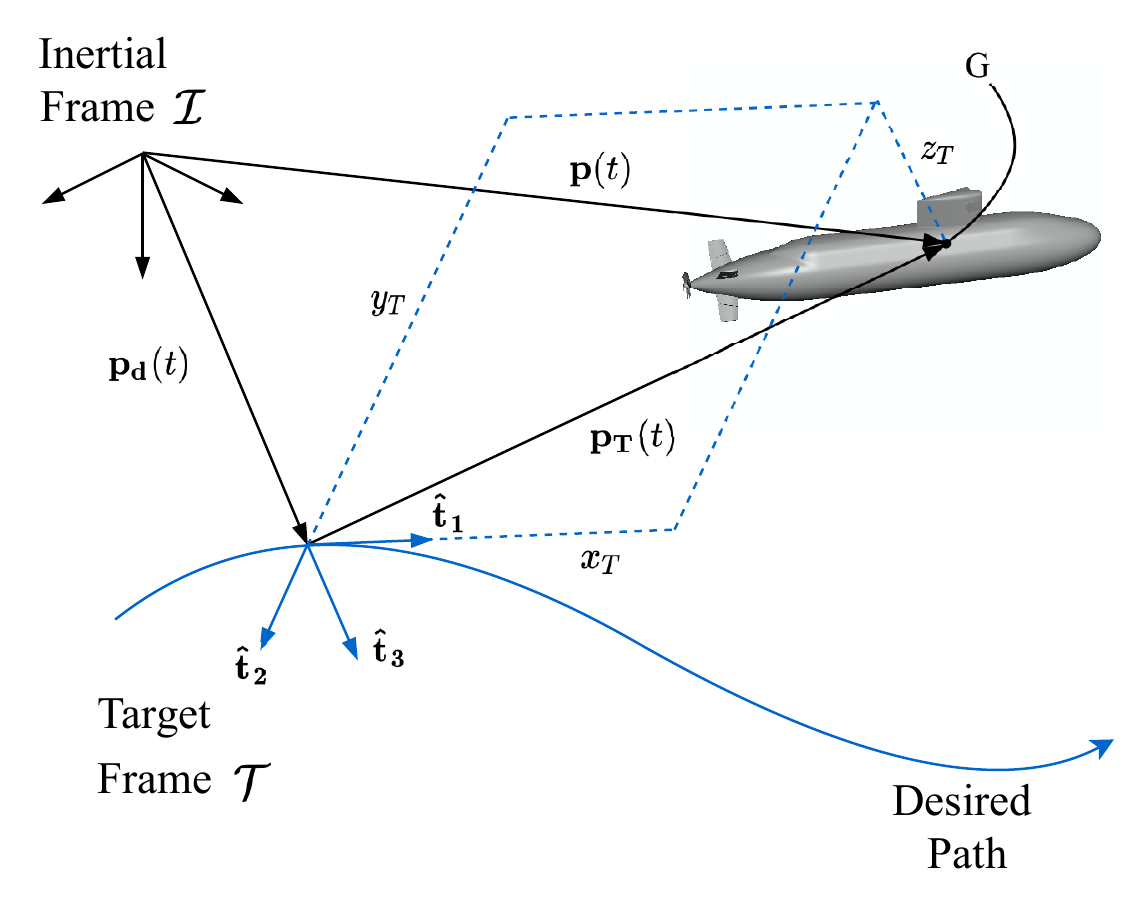}
    \caption{Geometry associated with the path-following problem, borrowed from \cite{rober2021three}}
    \label{fig:vector_summary}
\end{figure}

The path-following algorithm used to follow the generated trajectory draws from the high-level motion-control algorithm outlined in \cite{cichella2011geometric}, \cite{kaminer2017time}, \cite{cichella20113d}, \cite{cichella20133dmultirotor}. The geometric path $\bm{p_d}:[0,T_f] \to \mathbb{R}^3$ defined in an inertial frame, $\mathcal{I}$, is parametrized by virtual time, $\gamma:\mathds{R}^+\rightarrow[0,T_f]$, which allows the control law to move the virtual target along the path as a function of the vehicles state. A parallel transport frame \citep{kaminer2017time}, denoted as $\mathcal{T}$, defines the orientation of the virtual target, $\bm{p_d}(\gamma)$. The $\mathcal{T}$ frame’s orientation with respect to $\mathcal{I}$ is given by the rotation matrix $\bm{R}_T^I(\gamma) \triangleq [\bm{\hat t}_1(\gamma),\bm{\hat t}_2(\gamma),\bm{\hat t}_3(\gamma)]$, with $\bm{\hat t}_1(\cdot)$ being a unit vector tangent to the velocity of the path. $\bm{\hat t}_2(\gamma)$ and $\bm{\hat t}_3(\gamma)$ are orthonormal to $\bm{\hat t}_1(\cdot)$ and found considering curvature and torsion of the path at $\gamma$. The flow frame $\mathcal{W}$ is introduced with its origin, $\bm{p}$, at the vehicle center of mass. $\bm{R}_W^I \triangleq [\bm{\hat w}_1,\bm{\hat w}_2,\bm{\hat w}_3]$ gives the frames orientation with the x-axis being aligned to the vehicles velocity. The angular velocity of this frame with respect to $\mathcal{I}$ is denoted by $\bm{\omega}_W \triangleq [p,q,r]^{\top}$. Fig.~\ref{fig:vector_summary} provides a visual aid for the discussed frames and vectors.

\end{multicols}

\begin{multicols}{2}
\hspace{1mm}

With the defined reference frames, position error dynamics can be given in the $\mathcal{T}$ frame as, 
\begin{equation} \label{eq:positionerrordynamics}
    \dot{\bm{p_T}}   =
    \dot{\bm{R}}_{I}^{T}(\bm{p}-\bm{p_d}(\gamma)) + {\bm{R}}_{I}^{T} \dot{\bm{p}} - {\bm{R}}_{I}^{T} \dot{\bm{p}}_{\bm{d}}.
\end{equation}
An additional desired frame $\mathcal{D}$ is defined at the vehicles center of mass to help derive the path-following attitude error. The dynamics of this error can be written as
\begin{equation} \label{eq:attitudeerrordynamics}
	{\dot{\Psi}}(\bm{\tilde{R}}) = \bm{e}_{\bm{\tilde{R}}}^\top \left(\begin{bmatrix} \;q\; \\ r \end{bmatrix} - {\bm{\Pi_R}} \bm{\tilde{R}}^\top\left(\bm{R_{T}^{D}}{\bm{\omega}}_T +\bm{\omega}_{DT}^D\right)\right),
\end{equation}
where $\tilde{\bm{R}}$ is the orientation of the $\mathcal{D}$ frame with respect to $\mathcal{W}$, $\bm{\Pi_R} = \begin{bmatrix} 0 && 1 && 0 \\ 0 && 0 && 1\end{bmatrix}$, $\bm{\omega}_{DT}^D$ is the angular rate of frame $\mathcal{D}$ with respect to frame $\mathcal{T}$ resolved in $\mathcal{D}$, and 
\begin{equation*} \label{eq:eRdef}
\bm{e}_{\bm{\tilde{R}}}  \triangleq   \frac{1}{2}\bm{\Pi_R}\left(\left(\bm{\mathds{I}_3}-\bm{\Pi_R}^\top\bm{\Pi_R}\right)\bm{\tilde{R}}-\bm{\tilde{R}}^\top\left(\bm{\mathds{I}_3}-\bm{\Pi_R}^\top\bm{\Pi_R}\right)\right)^\vee, 
\end{equation*}
as shown in \cite{cichella2011geometric,lee2010geometric}.

To solve the defined path-following problem, the dynamics of virtual time are governed by,
\begin{equation} \label{eq:gammad}
	\dot{\gamma} = \frac{\left[{v}\bm{\hat{w}_1}+k_\gamma (\bm{p}-\bm{p_d}(\gamma))\right]^\top		\bm{\hat{t}}_{1}(\gamma)}{\left\Vert\bm{p_d'}(\gamma)\right\Vert} \, 
\end{equation} 
and the pitch- and yaw-rate commands are given by,
\begin{equation} \label{eq:qcomm}
 	\bm{\omega_c} \triangleq
 	\begin{bmatrix} \; q_c \; \\ \; r_c \; \end{bmatrix}
 	= \bm{\Pi_R} \bm{\tilde R}^\top \left( \bm{R}_{T}^{D}\bm{\omega}_T + \bm{\omega}_{DT}^{D}\right) - 2k_{\tilde{R}} \bm{e_{\tilde{R}}} \, .
\end{equation}
Here $k_\gamma > 0$ and $k_{\tilde{R}} >0$ are control gains. Assuming the vehicle is equipped with an inner-loop controller that provides tracking capabilities of feasible pitch- and yaw-rate commands, $q_c(t)$ and $r_c(t)$, which satisfy
\begin{equation*} \label{eq:commandsconstraints}
\left\Vert
\begin{bmatrix}
q_c(t) \\
r_c(t)
\end{bmatrix}
\right\Vert
\leq \omega_{c,\max} 
    % | q_W(t) - q_c(t) |\leq \delta_q \, , \quad |r_W(t) -r_c(t)| \leq \delta_r 
\, , \quad \forall t \geq 0
\end{equation*} 
for some $\omega_{c,\max}>0$, $$ \omega_{c, \max}-\omega_{T,\max}\dot{\gamma}-\sup_{t \geq 0} \left\Vert {\bm{\omega}}_{DT}^D \right\Vert > 0 $$ will be satisfied. Continuing, define
\begin{equation*} \label{eq:paramc}
\begin{split}
    & c < \min \left\{ \frac{1}{\sqrt{2}} , \frac{1}{2k_{\tilde R}} \left(\omega_{c, \max}-\omega_{T,\max}\dot{\gamma}-\sup_{t \geq 0} \left\Vert {\bm{\omega}}_{DT}^D \right\Vert \right) \right\} \\ & c_1 > 0
\end{split}
\end{equation*}
and
\begin{equation*} \label{eq:paramlambda}
    \lambda <  \frac{v_{\min}}{c_1^2 \sqrt{d^2+c^2c_1^2}} , \quad 0< \delta_\lambda < 1 \, . 
\end{equation*}
Assuming
\begin{equation*}
\left\Vert
\begin{bmatrix}
q_c(t) - q(t) \\
r_c(t) -r(t)
\end{bmatrix}
\right\Vert
< 2 \lambda \delta_\lambda c  
\, , \quad \forall t \geq 0 ,
\end{equation*}
there exist control parameters $d$, $k_\gamma$, and $k_{\tilde{R}}$ such that, for any initial state $\bm{e}_{\text{PF}}(0) \in \Omega_{\text{PF}}$, with
 \begin{equation} \label{eq:domainofattraction}
     \Omega_{\text{PF}} = \left\{ \bm{e}_{PF} : \Psi(\bm{\tilde R}) + \frac{1}{c_1^2} \Vert \bm{p_T} \Vert^2 \leq c^2 \right\}  ,
 \end{equation}
the rate of progression of the virtual time defined in equation \eqref{eq:gammad} and the angular-rate commands defined in \eqref{eq:qcomm} ensure that the path-following error $\bm{e}_{\text{PF}}(t)$ is locally uniformly ultimately bounded. More information on these bounds and this controller can be found in \cite{rober2021three}.

With the angular rate commands determined by the path-following controller, the inner loop controller ensures that those commands are properly carried out. Assuming that the submarine is already equipped with an autopilot which can follow input pitch- and yaw-rate reference signals by manipulating $\delta_V$ and $\delta_H$ as in equation \eqref{CommandsVH}, an $\mathcal{L}_1$ augmentation is formulated for it in order to improve performance. 

The autopilot-vehicle system is given as
\begin{equation}
    \bm{\mathcal{G}_p}(s)
    \begin{cases}
    \label{eq:autopilot}
        \dot{\bm{x}}(t) \!=\! \bm{A_{p}} \bm{x}(t) \!+\! \bm{B_{p}} (\bm{u_{ad}}(t)\!+\!\bm{f}(t,\bm{x}(t)))\\
        \bm{y}(t) \!=\! \bm{C_{p} x}(t),  \\
    \end{cases}
\end{equation}
where $\bm{u_{ad}}(t)$ is the input pitch- and yaw-rate reference signals, the output $\bm{y}(t)$ is the actual vehicle pitch- and yaw-rate rates and $\bm{f}(t,\bm{x}(t))$ is a time-varying function capturing system uncertainties and external disturbances. $\left\{\bm{A_{p}},\;\bm{B_{p}},\;\bm{C_{p}}\right\}$ is a controllable-observable triple describing the system.

A desired system,
 $
 \bm{M}(s) \triangleq \bm{C_{m}}\left(s\mathds{I}-\bm{A_{m}}\right)^{-1}\bm{B_{m}}
$,
is introduced as a design parameter of the $\mathcal{L}_1$ controller, specifying the desired pitch- and yaw-rate behavior. The system must be selected such that $\bm{A_m}$ is Hurwitz, $\bm{C_m}\bm{B_m}$ is nonsingular, and $\bm{M}(s)$ does not have a non-minimum-phase transmission zero. The dynamics of this desired system are given by
\begin{equation}
\label{eq:desired}
    \bm{y_m}(s) = \bm{M}(s) \bm{K_{g}} \bm{\omega_c}(s), \quad \bm{K_{g}} = -(\bm{C_{m}} \bm{A_{m}}^{-1} \bm{B_{m}})^{-1} .
\end{equation}

The adaptation law of this controller provides a discrete-time estimate $\bm{\hat{\sigma}_d}(t)$ for the unknown function $\bm{f}(t,\bm{x}(t))$ in equation \eqref{eq:autopilot}, and is given by
\begin{equation}
\label{eq:adaptation_law_}
    \bm{\hat{\sigma}_d}[i] = - \bm{\Phi}^{-1}(T_s)e^{\bm{\Lambda} \bm{A_{m}} \bm{\Lambda}^{-1}T_s}\bm{1}_{n_m2}(\bm{\hat{y}_d}[i]-\bm{y_d}[i])
\end{equation}
where
$
\bm{\hat{\sigma}_d}(t) = \bm{\hat{\sigma}_d}[i],\quad t \in [iT_s,(i+1)T_s),\quad i \in \mathds{Z}_{\geq 0}
$,
$y_d[i]=y(iT_s)$ is the sampled output, 
$$
\Phi (T_s) = \int_0^{T_s} e^{\Lambda A_m \Lambda^{-1} (T_s-\tau)} \Lambda d\tau ,
$$
and 
$$
\Lambda = 
\begin{bmatrix} 
C_m \\ D \sqrt{P} 
\end{bmatrix} ,
$$
where $P$ is matrix solution to $A_{m}^\top P+P A_{m}=-Q$ for a given positive definite matrix $Q$, $P = \sqrt{P}^\top \sqrt{P}$, and $D$ is a matrix that satisfies
$
D\left( C_m \left(\sqrt{P}\right)^{-1} \right)^{\top} = 0 .
$
With this estimation in place, a discrete-time output predictor can be formulated as
\begin{equation}
\label{eq:output_predictor}
\begin{split} 
    \bm{\hat{x}_d}[i+1] & = e^{\bm{A_{m}}T_s}\bm{\hat{x}_d}[i] \\ &  +\bm{A_{m}}^{-1}\left(e^{\bm{A_{m}} T_s}-\mathds{I}_{n_m}\right)\left(\bm{B_{m}} \bm{u_d}[i]+\bm{\hat{\sigma}_d}[i]\right), \\   \bm{\hat{y}_d}[i] & = \bm{C_{m}}\bm{\hat{x}_d}[i],\quad \bm{x_d}[0] = \bm{C_{m}}^{\dagger}\bm{y}_0
\end{split}
\end{equation}
which replicates the desired closed-loop dynamics of the system given by Equation \eqref{eq:desired}.

Finally, the control law giving the input $\bm{u_{ad}}$ is given as
\begin{equation*}
    \bm{u_{ad}}(t) = \bm{u_d}[i],\quad t \in [iT_s,(i+1)T_s),\quad i \in \mathds{Z}_{\geq 0}
\end{equation*}
where
\begin{equation}
\label{eq:control_law}
\begin{split}
    \bm{x_u}[i+1] & = e^{\bm{A_o}T_s}\bm{x_u}[i] \\ & +\bm{A_o}^{-1}\left(e^{\bm{A_o} T_s}-\mathds{I}_{n_o}\right)\left(\bm{B_o}e^{-\bm{A_{m}}T_s}\bm{\hat{\sigma}_d}[i]\right), \\
    \bm{u_d}[i] & = \bm{K_{g}}\bm{\omega_c}[i]-\bm{C_o}\bm{x_u}[i],\quad \bm{x_u}[0] = 0
\end{split}
\end{equation}
Here, the triple $\left\{\bm{A_{o}},\;\bm{B_{o}},\;\bm{C_{o}}\right\}$ is the minimal state-space realization of the transfer function
\begin{equation*}
\label{eq:filter}
    \bm{O}(s) = \bm{C}(s)\bm{M}^{-1}(s)\bm{C_{m}}\left(s\mathds{I}_{n_o}-\bm{A_{m}}\right)^{-1}
\end{equation*}
and $\bm{C}(s)$ is a strictly proper stable transfer function such that $\bm{C}(0) = \mathds{I}_2$. \cite{hovakimyan2010} and \cite{jafarnejadsani2018robust} should be consulted for discussion of the bounds of this controller and for information on parameter tuning.

\section*{RESULTS}

\subsection*{Validation of ROM}

Several canonical cases were considered to validate the model, including self-propulsion near the surface in calm water and in  waves, roll decay, controlled turning circles, horizontal and vertical zigzags. A comprehensive description of these results has been presented in \cite{kimthesis}, and we only present a summary of the vertical zigzag (VZZ) maneuvers here.

 The 10/10 VZZ maneuvers were completed at constant propeller rotational speed, using the same PID controller as the one used by \cite{carrica2021cfd} for full scale computations. four approach speeds were used: 3, 6 and 10 kts, the speeds used to develop the model, and an additional speed of 15 kts outside the range of the model. All maneuvers are started at self-propulsion and initiated with a negative deflection of control planes  to $-10^\circ$, maintained until a pitch angle of $10^\circ$ is reached, at which point the command is reversed to obtain the zigzag trajectory.
 
\begin{figure}[H]
  \centering
    \includegraphics[width=\columnwidth]{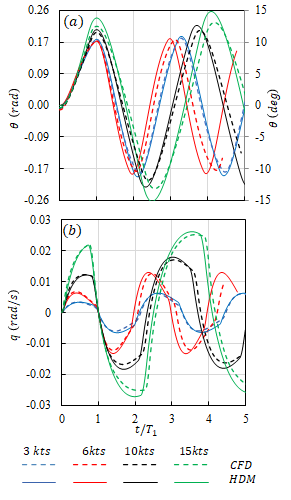}
    \caption{Pitch (top) and pitch rate (bottom) for 10/10 VZZ at different  speeds. Time non-dimensionalized by time to first positive pitch peak $T_1$.}
    \label{fig:VZZ_t}
\end{figure}

\begin{figure}[H]
  \centering
    \includegraphics[width=\columnwidth]{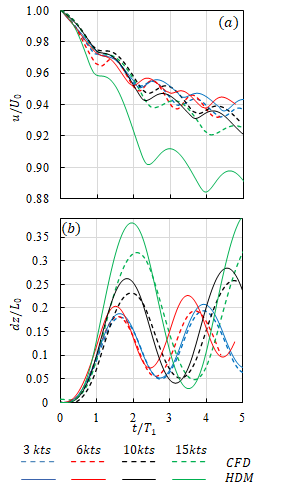}
    \caption{Forward speed (top) and depth change (bottom) for 10/10 VZZ at different speeds. Time non-dimensionalized by time to first positive pitch peak $T_1$.}
    \label{fig:VZZ_u}
\end{figure}

 The histories of motions as function of time normalized by the time  elapsed until the first positive peak in pitch, $T_1$, for each speed are shown in Figs.~\ref{fig:VZZ_t} and \ref{fig:VZZ_u}.
 

The overshoot angles and the zigzag period are in close match with the CFD simulations, with the largest error occurring at lower speeds. The pitch rate $q$ matches CFD particularly well, as shown in Fig.~\ref{fig:VZZ_t}(b), indicating that the control forces and moments are well predicted by the HDM. 

The prediction of vehicle speed loss is a key factor in the performance of the model, since all hydrodynamic forces depend on the speed. The good prediction of the speed shown in Fig.~\ref{fig:VZZ_u}(a) for velocities up to 10 kts indicates that the resistance and propeller thrust variation modeled from free running CFD simulation results are sufficiently accurate within the range of calibration of the model. The depth time histories presented in Fig.~\ref{fig:VZZ_u}(b) exhibit similar trends to the other motions, with good agreement with CFD but showing deterioration as the velocity increases. Note that even though much larger than at other speeds, the errors at 15 kts are still modest. 

\subsection*{Path following example}
An example of maneuvering following a pre-established trajectory near complex bathymetry is presented in this section. The geometry of the Scripps and La Jolla  canyons, as provided by \cite{barnard2010seamless} as a high resolution Digital Elevation Map (DEM), has been included in the simulation. The objective of this calculation is to demonstrate the ability of the controller to operate following a complex path both in the ROM and CFD. For the CFD case, the canyon geometry is included as an immersed boundary, therefore the interaction with the boundary is limited to a response to the pressure field. Moreover, since the simulations are set to follow a predetermined path, corrections to the vehicle trajectory are done to try to maintain the submarine on the path and do not account for corrections to avoid collisions. An additional simulation in CFD was conducted without considering the wall geometry which is a condition that more closely matches the simulation  using the ROM.

\begin{figure}[H]
  \centering
    \includegraphics[width=.95\columnwidth]{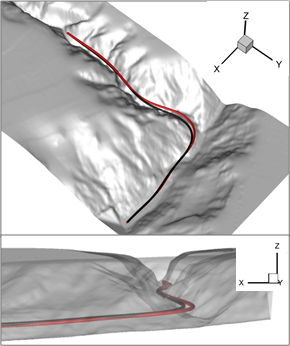}
    \includegraphics[width=.95\columnwidth]{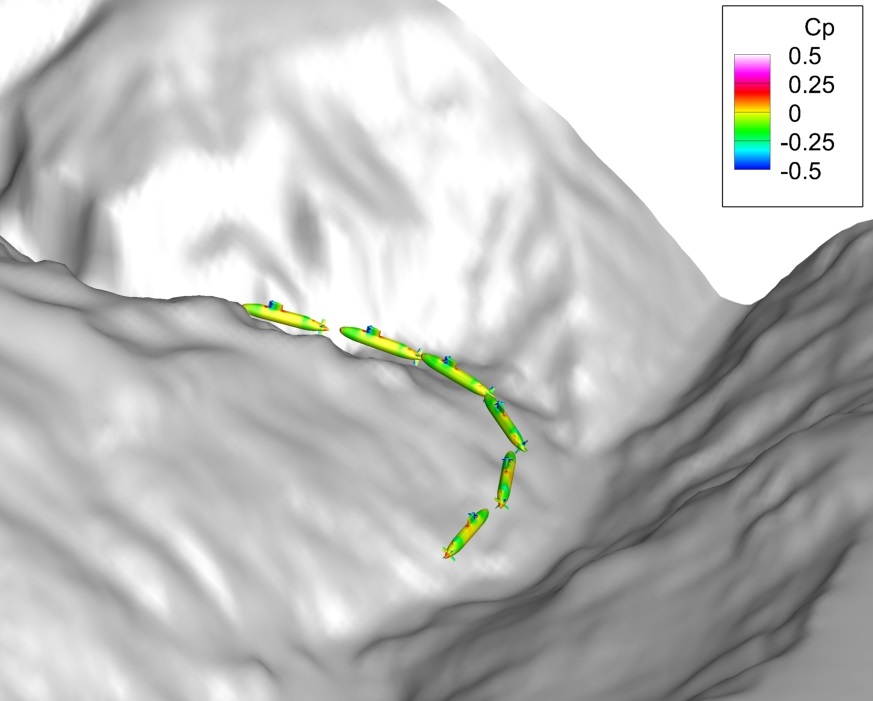}
    \caption{Top: Scripps and La Jolla canyons with prescribed path $\bm{p_d}$ (black) and actual path (red) for CFD simulation without considering walls; bottom: various BB2 position during maneuver simulated using immersed boundaries for the bathymetry. The latest position shown is immediately before collision with the wall. }
    \label{fig:canyon_geo}
\end{figure}

\begin{figure}[H]
  \centering
    \includegraphics[width=\columnwidth]{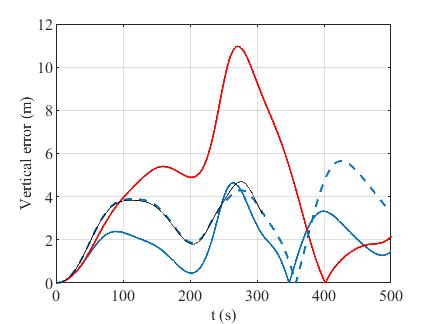}
    \includegraphics[width=\columnwidth]{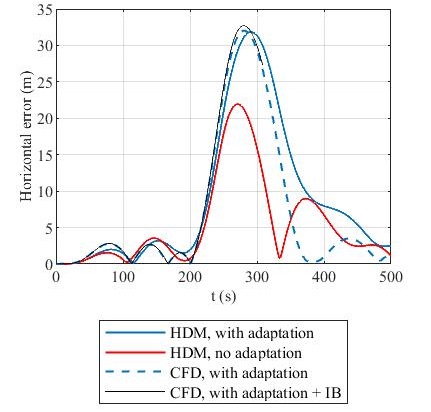}
    \caption{Error between target and actual path. Top: vertical error; bottom: horizontal error.}
    \label{fig:canyon_errors}
\end{figure}

Fig.~\ref{fig:canyon_geo} illustrates the bathymetry of the canyons and the target path. The path is set to approximately follow the thalweg (the line of lowest elevation at each cross-section within a valley) of the canyons maintaining a distance of 50 m between the CG and the bottom. The maneuver is completed at constant rps, and the corresponding speed is approximately 4 m/s in straight ahead conditions. The maneuver is completed after 500 s, which corresponds to a travelled distance approximately 1 mile from beginning to end location. The simulation considering the bathymetry via an immersed boundary stops at approximately 300~s, as the craft intercepts the wall. Both CFD simulations (with and without immersed boundaries) follow the same initial trajectory, indicating that the effect of the walls is negligible until the craft is in very close proximity to the surface. The collision occurs after completing the main turn of the maneuver, which coincides with the largest error between target and actual paths, as shown in Fig~\ref{fig:canyon_errors}.  Note also in Fig~\ref{fig:canyon_errors} that path errors are much larger in the horizontal. The controller used has been extensively tuned in the vertical to take advantage of the adaptive process, but less in the horizontal which can explain the differences observed. Note also that use of adaptation greatly reduces the vertical error, but slightly increases the horizontal error further indicating that more improvement of the controller is possible. A crucial result illustrated by Fig~\ref{fig:canyon_errors} is that the ROM produces comparable results to those from CFD, indicating that the ROM is capturing the craft's dynamics reasonably well. The location of extrema is consistently matched,  as well as their value for many of them.

\begin{figure}[H]
  \centering
    \includegraphics[width=\columnwidth]{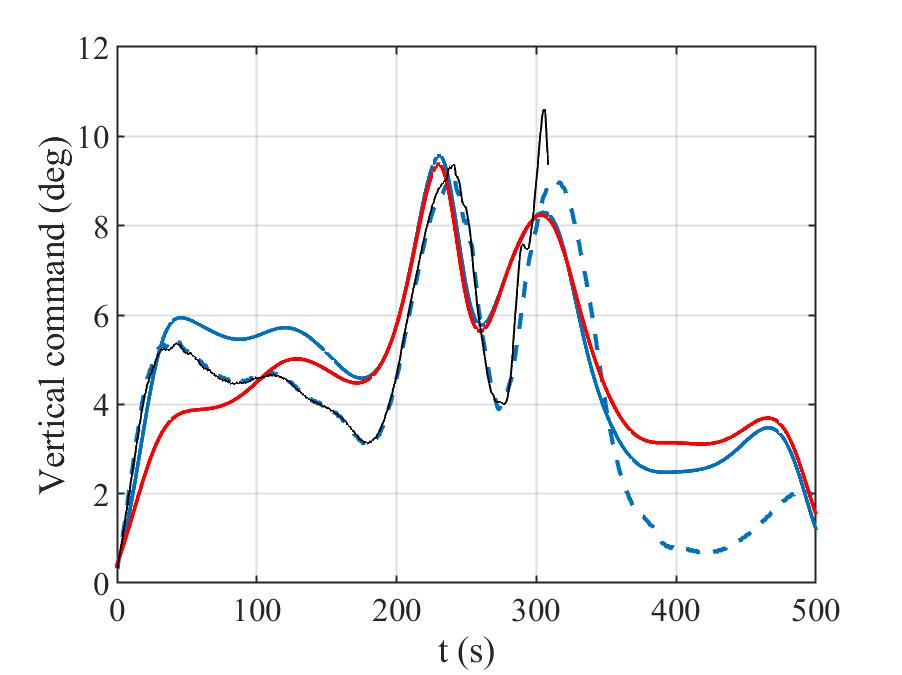}
    \includegraphics[width=\columnwidth]{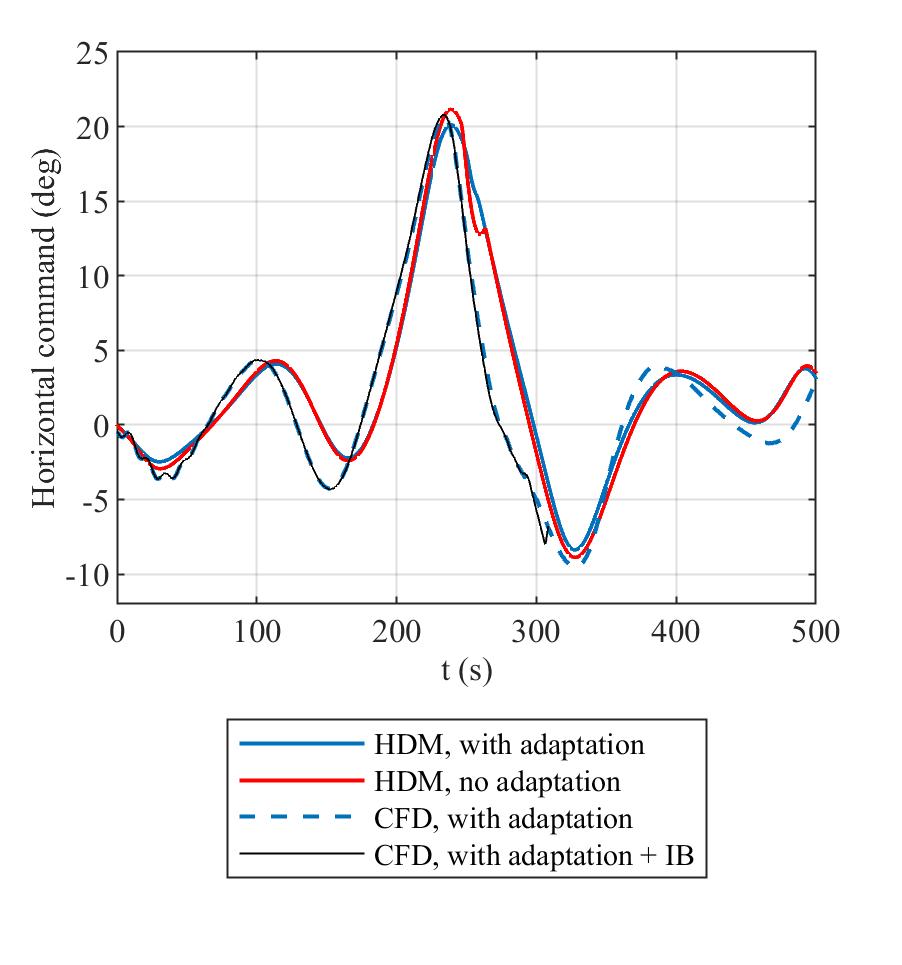}
    \caption{Control commands. Top: vertical; bottom: horizontal command.}
    \label{fig:canyon_commands}
\end{figure}

The trajectory of the vehicle is of course determined by the actuation of the control surfaces, and it is not surprising that the controller imposes similar command histories for the  two methods, as shown in Fig.~\ref{fig:canyon_commands}. Note that due to the combined effect of vertical and horizontal commands on the stern planes, the errors in tracking can be very different even though the commands are similar. This is particular noticeable when comparing results with and without adaptation for the ROM. 

\begin{figure}[H]
  \centering
    \includegraphics[width=\columnwidth]{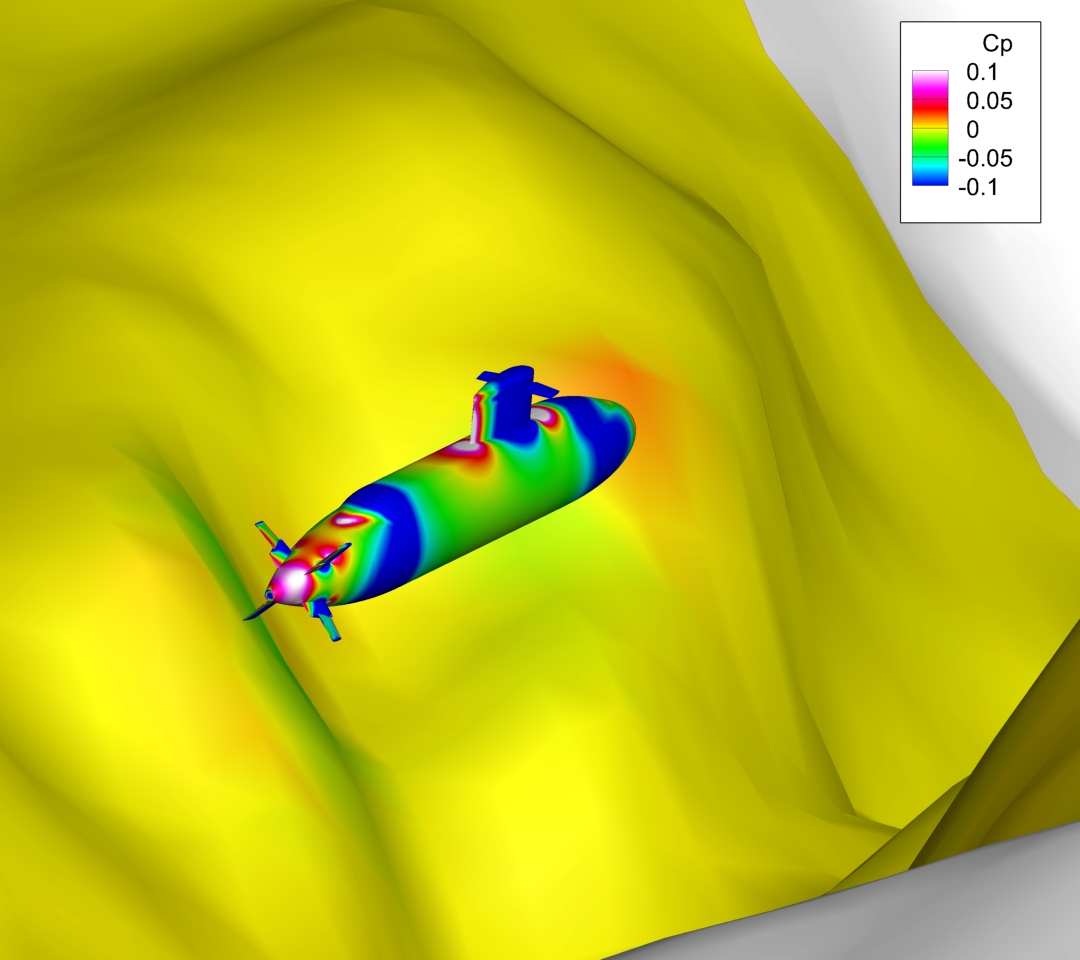}
    \includegraphics[width=\columnwidth]{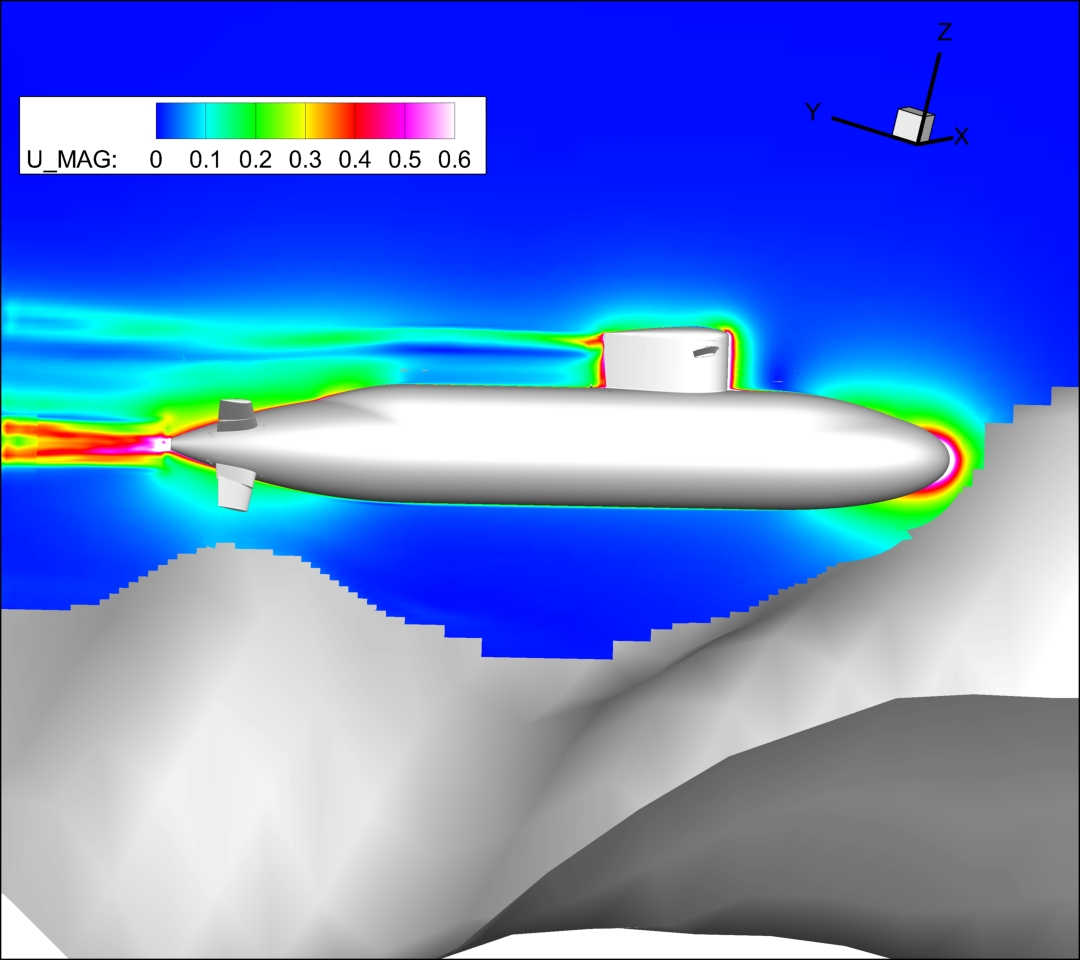}
    \caption{Flow field. Top: pressure distribution on the canyon wall (extrapolated from immersed boundary); bottom: dimensionless velocity magnitude ($U_0 = $ 4~m/s). Lower edge of flow field corresponds to the immersed boundary.}
    \label{fig:canyon_flow}
\end{figure}

The two CFD results show that the vertical command differs only noticeably immediately before the simulation considering an immersed boundary is terminated. This difference can only originate from an interaction with the immersed boundary. Examining the flow field, a weak interaction with the wall is captured, as shown in Fig.~\ref{fig:canyon_flow}. The different reaction of the controller does not alter significantly the trajectory of the craft and does not provide additional information regarding the interaction with the boundary. Further investigation of this interaction is granted, particularly with regards to the appropriate discretization needed for the background grid to accurately capture blocking effects with an immersed boundary, which is probably only captured for the current simulation in regions very near the vehicle surface, as it can be inferred by the very coarse discretization of the immersed boundary displayed in Fig.~\ref{fig:canyon_flow}.

\end{multicols}
\begin{multicols}{2}

\section*{SUMMARY AND CONCLUSIONS}

A reduced order model (ROM) of a generic submarine was presented. The model was developed to aide on the development of an adaptive controller, which is also briefly presented here. The hydrodynamic model developed is based on coefficients obtained exclusively from CFD simulations. The equations of the model, and some examples of the calculation of its coefficients was presented, however the complete set cannot be shown due to space constraints. The controller and the model performance was demonstrated using an example that requires the tracking of a complex path. Results from the ROM compare favorably with those from CFD. Wall effects were incorporated  as an immersed boundary only to the CFD case; the effect of the boundary appears very weak for the condition simulated, however this could be a result of the use of a grid that is  too coarse in the background and thus is unable to transmit the pressure effects unless the craft is extremely close to the bottom. Adding external forcing, such as a current or a wave field would also result in a stronger are more noticeable response form the immersed boundary, and such computations are part of our future work. 

Improvements of the hydrodynamic model are on-going. The dominant terms, partially shown here, such as hull drift, control surfaces drag and lift, and propulsion, have been validated, as well as many other terms obtained from rotating tests; however some secondary terms in Eqs.~\eqref{eq:fh_x} through \eqref{eq:fh_q} are still being evaluated. Even for the most dominant terms there are areas of improvement, from extension to even lower speeds, to the analysis of conditions that combine effects considered independent for the initial generation of the model.

An additional area of investigation for the ROM is the effect of boundaries, which is only incorporated in the current model considering the free surface. The example presented does not show strong interaction with the solid boundary in the CFD, however further investigation of this critical phenomenon is warranted.

%\columnbreak
\section*{ACKNOWLEDGEMENTS}
This research was supported by the Office of Naval Research, grant N00014-19-2106.

\end{multicols}
\begin{multicols}{2}
\bibliography{references}

\begin{thebibliography}{}

\bibitem[Barnard and Hoover, 2010]{barnard2010seamless}
Barnard, P.~L. and Hoover, D. (2010).
\newblock A seamless, high-resolution, coastal digital elevation model ({DEM})
  for {S}outhern {C}alifornia.
\newblock Technical report, US Geological Survey.

\bibitem[Carrica et~al., 2021a]{carrica2021cfd}
Carrica, P.~M., Kerkvliet, M., Quadvlieg, F., and Martin, J.~E. (2021a).
\newblock {CFD} simulations and experiments of a submarine in turn, zigzag, and
  surfacing maneuvers.
\newblock {\em Journal of Ship Research}, 65(04):293--308.

\bibitem[Carrica et~al., 2019]{carrica2019}
Carrica, P.~M., Kim, Y., and Martin, J.~E. (2019).
\newblock Near-surface self propulsion of a generic submarine in calm water and
  waves.
\newblock {\em Ocean Engineering}, 183:87--105.

\bibitem[Carrica et~al., 2021b]{carrica2021vertical}
Carrica, P.~M., Kim, Y., and Martin, J.~E. (2021b).
\newblock Vertical zigzag maneuver of a generic submarine.
\newblock {\em Ocean Engineering}, 219:108386.

\bibitem[Carrica et~al., 2007]{carrica2007ship}
Carrica, P.~M., Wilson, R.~V., Noack, R.~W., and Stern, F. (2007).
\newblock Ship motions using single-phase level set with dynamic overset grids.
\newblock {\em Computers \& Fluids}, 36(9):1415--1433.

\bibitem[Cichella et~al., 2013]{cichella20133dmultirotor}
Cichella, V., Choe, R., Mehdi, S.~B., Xargay, E., Hovakimyan, N., Kaminer, I.,
  and Dobrokhodov, V. (2013).
\newblock A 3{D} path-following approach for a multirotor {UAV} on {SO} (3).
\newblock {\em IFAC Proceedings Volumes}, 46(30):13--18.

\bibitem[Cichella et~al., 2011a]{cichella2011geometric}
Cichella, V., Kaminer, I., Dobrokhodov, V., Xargay, E., Hovakimyan, N., and
  Pascoal, A. (2011a).
\newblock Geometric 3{D} path-following control for a fixed-wing {UAV} on {SO}
  (3).
\newblock In {\em AIAA Guidance, Navigation, and Control Conference}, page
  6415.

\bibitem[Cichella et~al., 2019]{cichella2019optimal}
Cichella, V., Kaminer, I., Walton, C., Hovakimyan, N., and Pascoal, A.~M.
  (2019).
\newblock Consistent approximation of optimal control problems using
  {B}ernstein polynomials.
\newblock In {\em 2019 IEEE 58th Conference on Decision and Control (CDC)},
  pages 4292--4297.

\bibitem[Cichella et~al., 2011b]{cichella20113d}
Cichella, V., Naldi, R., Dobrokhodov, V., Kaminer, I., and Marconi, L. (2011b).
\newblock On 3{D} path following control of a ducted-fan {UAV} on {SO} (3).
\newblock In {\em 2011 50th IEEE Conference on Decision and Control and
  European Control Conference}, pages 3578--3583. IEEE.

\bibitem[Evans and Nahon, 2004]{evans2004dynamics}
Evans, J. and Nahon, M. (2004).
\newblock Dynamics modeling and performance evaluation of an autonomous
  underwater vehicle.
\newblock {\em Ocean Engineering}, 31(14-15):1835--1858.

\bibitem[Fang et~al., 2006]{fang2006wave}
Fang, M.-C., Chang, P.-E., and Luo, J.-H. (2006).
\newblock Wave effects on ascending and descending motions of the autonomous
  underwater vehicle.
\newblock {\em Ocean Engineering}, 33(14-15):1972--1999.

\bibitem[Feldman, 1995]{feldman1995method}
Feldman, J.~P. (1995).
\newblock Method of performing captive-model experiments to predict the
  stability and control characteristics of submarines.
\newblock Technical report, NSWC Carderock, Bethesda MD.

\bibitem[Gertler and Hagen, 1967]{gertler1967standard}
Gertler, M. and Hagen, G.~R. (1967).
\newblock Standard equations of motion for submarine simulation.
\newblock Technical report, David W Taylor Naval Ship Research and Development
  Center, Bethesda MD.

\bibitem[Hovakimyan and Cao, 2010]{hovakimyan2010}
Hovakimyan, N. and Cao, C. (2010).
\newblock {\em L1 adaptive control theory: Guaranteed robustness with fast
  adaptation}.
\newblock Society for Industrial and Applied Mathematics.

\bibitem[Jafarnejadsani, 2018]{jafarnejadsani2018robust}
Jafarnejadsani, H. (2018).
\newblock {\em Robust adaptive sampled-data control design for MIMO systems:
  Applications in cyber-physical security}.
\newblock PhD thesis, University of Illinois at Urbana-Champaign.

\bibitem[Jin and Er, 2020]{jin2020dynamic}
Jin, X. and Er, M.~J. (2020).
\newblock Dynamic collision avoidance scheme for unmanned surface vehicles
  under complex shallow sea environments.
\newblock {\em Ocean Engineering}, 218:108102.

\bibitem[Joubert, 2004]{joubert2004some}
Joubert, P.~N. (2004).
\newblock Some aspects of submarine design. {P}art 1. {H}ydrodynamics.
\newblock Technical report, Defence Science and Technology Organisation,
  Victoria, Australia.

\bibitem[Joubert, 2006]{joubert2006some}
Joubert, P.~N. (2006).
\newblock Some aspects of submarine design. {P}art 2. {S}hape of a submarine
  2026.
\newblock Technical report, Defence Science and Technology Organisation,
  Victoria, Australia.

\bibitem[Kaminer et~al., 2017]{kaminer2017time}
Kaminer, I., Pascoal, A.~M., Xargay, E., Hovakimyan, N., Cichella, V., and
  Dobrokhodov, V. (2017).
\newblock {\em Time-Critical cooperative control of autonomous air vehicles}.
\newblock Butterworth-Heinemann.

\bibitem[Kim, 2021]{kimthesis}
Kim, Y. (2021).
\newblock {\em Development and validation of hydrodynamic model for near free
  surface maneuvers of {BB2} {J}oubert generic submarine}.
\newblock PhD thesis, The University of Iowa.

\bibitem[Lee et~al., 2010]{lee2010geometric}
Lee, T., Leok, M., and McClamroch, N.~H. (2010).
\newblock Geometric tracking control of a quadrotor {UAV} on {SE} (3).
\newblock In {\em 49th IEEE Conference on Decision and Control (CDC)}, pages
  5420--5425. IEEE.

\bibitem[Li et~al., 2020]{li2020modeling}
Li, J., Yuan, B., and Carrica, P.~M. (2020).
\newblock Modeling bubble entrainment and transport for ship wakes: progress
  using hybrid {RANS}/{LES} methods.
\newblock {\em Journal of Ship Research}, 64(04):328--345.

\bibitem[Menter, 1994]{menter1994two}
Menter, F.~R. (1994).
\newblock Two-equation eddy-viscosity turbulence models for engineering
  applications.
\newblock {\em AIAA journal}, 32(8):1598--1605.

\bibitem[Noack et~al., 2009]{noack2009suggar}
Noack, R., Boger, D., Kunz, R., and Carrica, P.~M. (2009).
\newblock Suggar++: An improved general overset grid assembly capability.
\newblock In {\em 19th AIAA Computational Fluid Dynamics}, page 3992.

\bibitem[Overpelt et~al., 2015]{overpelt2015free}
Overpelt, B., Nienhuis, B., Anderson, B., et~al. (2015).
\newblock Free running manoeuvring model tests on a modern generic {SSK} class
  submarine ({BB2}).
\newblock In {\em Pacific International Maritime Conference}, pages 1--14.

\bibitem[Rober et~al., 2021]{rober2021three}
Rober, N., Cichella, V., Martin, J.~E., Kim, Y., and Carrica, P.~M. (2021).
\newblock Three-dimensional path-following control for an underwater vehicle.
\newblock {\em Journal of Guidance, Control, and Dynamics}, pages 1--11.

\end{thebibliography}
\end{multicols}
\end{document}